\documentclass[12pt,a4paper]{article}
\usepackage[latin1]{inputenc}
\usepackage[T1]{fontenc}
\usepackage{amsmath}
\usepackage{amsfonts}
\usepackage{amssymb}
\usepackage[english]{babel}
\usepackage[dvips]{graphicx,color}
\title{Thermal boundary conditions as constraints}
\author{C.~D.~Fosco$^{a}$\\
A.~P.~C.~Malbouisson$^{b}$\\ and I.~Roditi$^{b}$\\ {\normalsize\it
$^a$Centro At\'omico Bariloche and Instituto Balseiro}\\
{\normalsize\it Comisi\'on Nacional de Energ\'\i a At\'omica} \\
{\normalsize\it 8400 Bariloche, Argentina.}\\ {\normalsize\it
$^b$Centro Brasileiro de Pesquisas F\'isicas - CBPF/MCT}\\
{\normalsize\it Rua Dr. Xavier Sigaud, 150, 22290-180 Rio de Janeiro,
RJ, Brazil}}
\begin{document}
\newcommand*{\be}{\begin{equation}}
\newcommand*{\ee}{\end{equation}}
\newcommand*{\beq}{\begin{eqnarray}}
\newcommand*{\eeq}{\end{eqnarray}}
\def\d{{\rm d}}
\maketitle
\begin{abstract}
\noindent We introduce the boundary conditions corresponding to the
imaginary-time (Matsubara) formalism for the finite-temperature partition
function in $d+1$ dimensions as {\em constraints\/} in the path integral
for the vacuum amplitude (the zero-temperature partition function).
We implement those constraints by using Lagrange multipliers, which are 
static fields, two of them associated to each physical degree of
freedom.
After integrating out the original, physical fields, we obtain an
effective representation for the partition function, depending only on the
Lagrange multipliers.  The resulting functional integral has the appealing
property of involving only $d$-dimensional, {\em time independent \/}
fields, looking like a non local version of the classical partition
function. 
We analyze the main properties of this novel representation for the
partition function, developing the formalism within the context of two
concrete examples: the real scalar and Dirac fields.  
\end{abstract}
\section{Introduction}\label{sec:intro}
 About fifty years ago, in a pioneering work~\cite{mats1}, the theoretical
foundations for a systematic treatment of Quantum Field Theory (QFT) at
finite temperature ($T>0$), were laid down.  That approach, the now called
Matsubara (or imaginary time) formalism has been very successful indeed in
allowing for the evaluation of thermal effects in QFT, both in the High
Energy~\footnote{See, for example~\cite{kap1,belac1}.} and Condensed Matter Physics realms.

Another landmark in the subsequent development of that formalism was
introduced in~\cite{ume1}, where Matsubara's work was extended to {\em
relativistic\/} QFT. The crucial discovery of the time-periodicity
(antiperiodicity) conditions for the bosonic (fermionic) Greens's functions
opened the door to a quite natural (and fruitful) extension of many
concepts and ideas originally introduced at $T=0$, to the finite
temperature context. To name just a few examples: the thermal Ward-Takahashi
relations, Goldstone's theorem, the Kubo-Martin-Schwinger (KMS) relations
and the renormalization group at $T>0$. Besides, the notion of Abelian and
non Abelian gauge fields, with all its consequences for particle
physics~\cite{kap1,belac1,le1} could also be set up in the finite
temperature context in a quite natural way.

In the path-integral framework for the imaginary time formalism, thermal
effects depend strongly on the fact that the imaginary-time
coordinate is compactified.  That is, fields corresponding to a
$d+1$-dimensional theory, must be defined on $S^1 \times R^{d}$, where the
radius of $S^{1}$ is proportional to the inverse temperature. This fact
has, of course, many important consequences for the construction of
topologically non-trivial solutions~\cite{Gross:1980br} since they naturally
depend upon the global properties of the spacetime manifold (unlike, for
example, the UV structure of the theory). On the other hand, the
topologically trivial sector is also affected by the fact that the
frequencies become {\em discrete\/} variables (the Matsubara frequencies).

A feature of the Matsubara formalism (shared by the real-time formulation)
is that a time dependence for the fields is unavoidable, even when one
limits oneself to the calculation of static thermal properties.  With the
aim of constructing a new representation where only static fields are
involved, in this paper, we introduce an alternative way of dealing with
finite-temperature QFT. It is inspired by a recent paper in which a
constrained functional integral approach is used to implement the effect of
fluctuating boundaries in the Casimir effect~\cite{kardar}. In the present
context, this allows one to introduce the periodicity conditions by means
of Lagrange multipliers (which are $d$-dimensional when the field lives in 
$d+1$ dimensions).  Then the original fields can be integrated, what leaves a
functional depending only on the $d$-dimensional Lagrange multipliers.
Besides the economy due to the `dimensional reduction', there are also some
new interesting properties, as we shall show in the different sections of
this article.

The structure of this paper is as follows: in section~\ref{sec:harmonic},
we introduce the constraint formalism to deal with $T>0$, for the simplest
possible example, i.e., the harmonic oscillator. Equipped with these
results, the case of the real scalar field is dealt with in
section~\ref{sec:real}, both for the free and the self-interacting cases.  In
section~\ref{sec:Dirac} we consider the Dirac field, and
in~\ref{sec:conclu}, we present our conclusions.
\section{Harmonic oscillator}\label{sec:harmonic}
We shall introduce the method by considering the simplest (yet non trivial)
example, the harmonic oscillator. In the subsequent sections, the same
approach shall be extended, in a quite natural way, to interacting QFT's.

Let us begin by writing the phase-space path integral~\cite{zinn} for
${\mathcal Z}_0$, the {\em vacuum\/} functional for the harmonic
oscillator:
\begin{equation}\label{eq:defz0}
{\mathcal Z}_0 \;=\; \int \, {\mathcal D}p \,{\mathcal D}q \; 
\exp \big\{-S_0[q(\tau),p(\tau)] \big\} \;,
\end{equation}
where $S_0$ is the imaginary time first-order action:
\begin{equation}\label{eq:defs0h}
S_0[q(\tau),p(\tau)]\,=\, 
\int_{-\infty}^{+\infty} d\tau 
\big[ - i p(\tau) \dot{q}(\tau) + {\mathcal H}_0\big( p(\tau),q(\tau)
\big)\big] \Big\}
\end{equation}
with ${\mathcal H}_0$ denoting the classical Hamiltonian,
\begin{equation}
{\mathcal H}_0(p,q) \;=\; \frac{1}{2} \big(p^2 \,+\, \omega^2 q^2\big)
\end{equation}
(for simplicity, we have assumed that the oscillator has unit mass).
We have introduced in (\ref{eq:defz0}) the formal expression for the
phase-space path integral integration measure:
\begin{equation}
{\mathcal D}p \,{\mathcal D}q \;\equiv\; \prod_{-\infty < \tau <\infty}
\frac{dp(\tau) dq(\tau)}{2\pi} \;,
\end{equation}
whose precise definition may, of course, be given in terms of a
discrete approximation to the path integral~\footnote{The reason why the phase
space path integral (rather than the more usual configuration space one) is
used, will become clear at the end of this section.}.

On the other hand, in the Matsubara formalism, ${\mathcal Z}_0(\beta)$, the
{\em thermal\/} partition function (for the same system) can be written as:
\begin{equation}\label{eq:defzh}
{\mathcal Z}_0(\beta) \;=\; 
\int_{\rm periodic}\, {\mathcal D} p \, {\mathcal D} q \; 
e^{\int_0^\beta d\tau \big[ i p \dot{q} 
\,-\,\frac{1}{2} (p^2 + \omega^2 q^2 )\big]} \;,
\end{equation}
where the integration is taken over periodic paths \mbox{$q(\beta)
= q(0)$}, \mbox{$p(\beta)= p(0)$}~\footnote{This symmetric form of
the boundary conditions in the path integral for ${\mathcal Z}_0(\beta)$ is
explained in~\cite{teitelboim}.} and we have used units such that $\beta =
\frac{1}{T}$, i.e., $k_B \equiv 1$.

To simplify matters, we shall measure the energies with respect to the
vacuum (zero energy).  For the partition function, this means that
\mbox{${\mathcal Z}_0(\infty) \equiv 1$}; hence, we shall absorb
temperature-independent factors by introducing a suitably chosen normalization 
constant in ${\mathcal Z}_0(\beta)$.

To reproduce ${\mathcal Z}_0(\beta)$, we shall follow a quite
straightforward procedure: we will impose the periodicity conditions for
the phase space variables as constraints for the integration variables in
${\mathcal Z}_0$.  To that end, we write: 
\begin{equation}\label{eq:zb1}
{\mathcal Z}_0(\beta) \,=\, {\mathcal N} \, \int {\mathcal D} p \, {\mathcal D}
q \;\delta\big(q(\beta)-q(0)\big) \, \delta\big(p(\beta)- p(0)\big)
\, e^{- S_0[q(\tau),p(\tau)]} \;,
\end{equation}
where ${\mathcal N}$ is a constant, introduced to enforce the condition 
\mbox{${\mathcal Z}_0(\infty) \equiv 1$}, and $S_0$ is the action defined
in (\ref{eq:defs0h}). 

The imaginary-time $\tau$ runs from $-\infty$ to $+\infty$, while the
constraints identify just the fields at $0$ and $\beta$. Thus one should
expect the emergence of a ${\mathcal Z}_0(\infty)$ factor in
(\ref{eq:zb1}), due to the integration of fields outside of the interval
$[0,\beta]$. That factor is, however, $\beta$-independent, and shall therefore
be absorbed by ${\mathcal N}$. 

To proceed, we exponentiate the two $\delta$-functions, by means of (two)
Lagrange multipliers $\xi_1$ and $\xi_2$ (they are just real variables in
this case); we are thus lead to a new expression for ${\mathcal
Z}_0(\beta)$, which may be written as follows:
\begin{eqnarray}\label{eq:aux1}
{\mathcal Z}_0(\beta) &=& {\mathcal N} \, \int {\mathcal D} p \, {\mathcal D} q 
\int_{-\infty}^{\infty} \frac{d\xi_1}{2\pi} \int_{-\infty}^{\infty}
\frac{d\xi_2}{2\pi} \nonumber\\
&\times& e^{i\int_{\infty}^{+\infty} d\tau \Big[
q(\tau) \, \xi_1 \,\big( \delta(\tau-\beta) - \delta(\tau) \big)
\,+\, p(\tau)\, \xi_2 \, \big( \delta(\tau-\beta) - \delta(\tau) \big)\Big]}
\nonumber\\
&\times& e^{\int_{-\infty}^\infty d\tau \big[ i p \dot{q} 
\,-\,\frac{1}{2} (p^2 + \omega^2 q^2 )\big]} \;.
\end{eqnarray}
The integral over $p$ and $q$ (which we shall perform first) is obviously
Gaussian. Indeed, interchanging the order of integration, we see that
(\ref{eq:aux1}) may be rewritten as follows:
\begin{eqnarray}\label{eq:zb2}
{\mathcal Z}_0(\beta) &=& {\mathcal N} \,\int_{-\infty}^{\infty}
\frac{d\xi_1}{2\pi} \int_{-\infty}^{\infty}\frac{d\xi_2}{2\pi}
\int {\mathcal D} Q 
\nonumber\\
&\times& e^{- \frac{1}{2} \int_{\infty}^{+\infty} d\tau \, Q_a(\tau) 
\widehat{{\mathcal K}}_{ab} Q_a(\tau) \,+\,i\, \int_{-\infty}^\infty d\tau
j_a(\tau) Q_a(\tau) } \;, 
\end{eqnarray}
where we have introduced  \mbox{$Q \,=\, (Q_a)$} ($a=1,\, 2$) for the
canonical coordinates, such that $Q_1 \equiv q$ and $Q_2 \equiv p$. Also,
\mbox{$j_a(\tau) \equiv \xi_a \,( \delta(\tau-\beta) - \delta(\tau))$}, and
$\widehat{\mathcal K}_{ab}$ are the elements of the $2 \times 2$ operator 
matrix $\widehat{\mathcal K}$, given by:
\begin{equation}\label{eq:defk}
\widehat{\mathcal K} \;=\; 
\left( 
\begin{array}{cc}
\omega^2              & i \frac{d}{d\tau} \\
- i \frac{d}{d\tau}     & 1 
\end{array}
\right) \;.
\end{equation}
The result of integrating out $Q_a$, may be written as follows:
\begin{equation}
{\mathcal Z}_0(\beta) \;=\; {\mathcal N}\, 2\pi \, 
\big(\det\widehat{\mathcal K}\big)^{-\frac{1}{2}} \;
\int \frac{d^2\xi}{(2\pi)^2} \,\, e^{- \frac{1}{2} \xi_a M_{ab} \xi_b } \;,
\end{equation}
with
\begin{equation}\label{eq:defm}
M \;\equiv\; 2 \, \Omega(0) \,-\, \Omega(\beta) \,-\, \Omega(-\beta) \;, 
\end{equation}
where $\Omega(\tau)$ denotes the inverse of the operator ${\mathcal K}$
of (\ref{eq:defk}); its explicit form may be easily found to be:
\begin{equation}\label{eq:defom}
\Omega(\tau)\;\equiv\;
\left( 
\begin{array}{cc}
\frac{1}{2\omega} & \frac{i}{2} {\rm sgn}(\tau) \\
-\frac{i}{2} {\rm sgn}(\tau) & \frac{\omega}{2} 
\end{array}
\right) \; e^{-\omega |\tau|} 
\end{equation}
(${\rm sgn} \equiv$ sign function). Equation (\ref{eq:defom}) can be used
in (\ref{eq:defm}), to write:
\begin{equation}\label{eq:m1}
M\;=\;
\left( 
\begin{array}{cc}
\omega^{-1} & 0 \\
0 & \omega 
\end{array}
\right) \; (n_B(\omega) + 1)^{-1} \;,
\end{equation}
where 
\begin{equation}
n_B(\omega) \equiv (e^{\beta \omega} - 1)^{-1}\
\end{equation}
is the Bose-Einstein distribution function.

The resulting integral over the $\xi_a$ variables becomes:
\begin{equation}\label{eq:zosc}
{\mathcal Z}_B \;=\; 
\int \frac{d^2\xi}{2\pi} \,
e^{- \frac{\omega^{-1} \, \xi_1^2 \,+\; 
\omega \; \xi_2^2}{2 [ n_B(\omega) + 1 ]}} \;,
\end{equation}
where we have set:
\begin{equation}
{\mathcal N} \;\equiv\; \big(\det\widehat{\mathcal K}\big)^{\frac{1}{2}}\;,
\end{equation}
by the normalization convention. 
The resulting integral is over the two (real) variables $\xi_a$, which are
$0$-dimensional fields, one dimension less than the $0+1$ dimensional original theory. 

It is interesting to see how this integral may be compared with the one
corresponding to classical statistical mechanics. To that end, we evaluate
the partition function in the classical (high-temperature) limit.
In that limit we approximate the integrand and ${\mathcal Z}_0(\beta)$ becomes:
\begin{equation}\label{eq:zosc2}
{\mathcal Z}_0(\beta) \;\simeq\;\int \frac{d^2\xi}{2\pi} \,
e^{- \beta H(\xi_1,\xi_2)} \;\;\; (\beta << 1) \;,
\end{equation}
where:
\begin{equation}
H(\xi_1,\xi_2) \;\equiv\; \frac{1}{2} \big( \xi_1^2 \,+\, \omega^2 \xi_2^2 
\big)\;.
\end{equation}
We see that (\ref{eq:zosc2}) corresponds exactly to the
classical partition function for a harmonic oscillator, when the identifications:
$\xi_1=p$ (classical momentum), and $\xi_2=q$ (classical coordinate) are
made
\begin{equation}\label{eq:zosc3}
{\mathcal Z}_0(\beta) \;\simeq\;\int \frac{dp dq}{2\pi} \,
e^{- \beta \frac{1}{2} \big( p^2 \,+\, \omega^2 q^2\big)}
\;\;\; (\beta << 1) \;.
\end{equation}

On the other hand, had the exact form of the integral been kept (no
approximation), we could still have written an expression similar to the
classical partition function, albeit with an `effective Hamiltonian'
$H_{eff}(\xi_1,\xi_2)$:
\begin{equation}\label{eq:zosc4}
{\mathcal Z}_0(\beta) \;=\;\int \frac{d^2\xi}{2\pi} \,
e^{- \beta H_{eff}(\xi_1,\xi_2)} \;,
\end{equation}
where:
\begin{equation}
H_{eff}(\xi_1,\xi_2) \;\equiv\; \frac{1}{2 \beta} \,
\big( n_B(\omega) + 1 \big)^{-1}\, \big(\omega^{-1} \, \xi_1^2 \,+\; 
\omega \; \xi_2^2 \big) \;.
\end{equation}
This shows that the quantum partition function may also be written as a
classical one, by using a $\beta$-dependent Hamiltonian, which of course
tends to its classical counterpart in the high-temperature limit.  This
representation is also valid for interacting theories, as shown
in~\ref{ssec:self}.

By integrating out the Lagrange multipliers in the (exact)
expression for the partition function (\ref{eq:zosc}), we obtain:
\begin{equation}\label{eq:free}
{\mathcal Z}_0(\beta) \;=\; n_B(\omega) \,+\, 1 \;=\; 
\frac{1}{1\,-\, e^{-\beta \omega}} \;.
\end{equation}
which is the correct result.

We emphasize an important fact that has emerged from an analysis of the
classical (high-temperature) limit in this case, namely, the Lagrange
multipliers have a physical interpretation in the classical limit: the one
associated to the periodicity condition for $q$ plays the role of the
classical momentum, while the one corresponding to the periodicity for the
momentum becomes a generalization of the classical coordinate. The same
interpretation might also be retained far from the classical limit, but then
the Hamiltonian departs from the classical one, receiving quantum corrections. 

To conclude this section, we mention the fact that the (apparently) simpler
procedure of working in terms of the {\em configuration space\/}
path-integral yields an {\em incorrect\/} answer. Indeed, had one used that
formulation, one should have needed to impose the periodicity constraint
just for $q(\tau)$:
\begin{equation}
{\mathcal Z}_0 (\beta) \;=\; {\mathcal N} \,\int\, {\mathcal D}q
\; \delta\big( q(\beta) - q(0)\big) \,\times \,
e^{-\frac{1}{2} \int_{-\infty}^{+\infty} d\tau \big(\dot{q}^2 +
\omega^2 q^2 \big)}\;.
\end{equation}
The $\delta$-function could then have been easily exponentiated by using
{\em just one\/} Lagrange multiplier, $\chi$:
\begin{equation}
{\mathcal Z}_0(\beta) \;=\; {\mathcal N} \, \int\, {\mathcal D}q \,
\int_{-\infty}^{+\infty} \frac{d\chi}{2\pi} \,
e^{ i \chi \,\big(  q(\beta) - q(0)\big)}
\, e^{ - \frac{1}{2} \int_{-\infty}^{+\infty} d\tau
\big(\dot{q}^2 + \omega^2 q^2\big)} \;.
\end{equation}
Performing the Gaussian integral over $q(\tau)$ would have lead to:
\begin{equation}
{\mathcal Z}_0(\beta) \;=\; {\mathcal N} \, {\mathcal Z}_0(\infty) \,
\int_{-\infty}^{+\infty} d\chi \; e^{ - Q(\chi)} \;,
\end{equation}
where:
\begin{equation}
Q(\chi) \;\equiv\; \omega^{-1} [n_B(\omega) + 1]^{-1} \chi^2 \,.
\end{equation}
Choosing ${\mathcal N}^{-1} \equiv {\mathcal Z}_0(\infty)$, and 
integrating out $\chi$, we would have seen that
\begin{equation}
{\mathcal Z}_0(\beta) \;=\;  \frac{1}{\sqrt{1 - e^{-\beta \omega}}} \;\;
\;\;(wrong)
\end{equation}
which is of course {\em incorrect\/}, due to the power $1/2$ (we have neglected
$\beta$-independent factors). 
The reason why this formulation yields an incorrect answer may be traced back
to the fact that it corresponds to a different constraint. Indeed, the
previous, phase space treatment, amounts to enforcing periodicity for the
coordinates and their momenta (essentially, their time derivatives). Thus, it
corresponds to requiring the periodicity for $q(\tau)$, and also the time
independence of that constraint.

\section{Real scalar field}\label{sec:real}
\subsection{Free partition function}
In what is perhaps the most straightforward extension to QFT of the previous
section's results, we shall consider here a real scalar field $\varphi$ in $d+1$
(Euclidean) dimensions: $\varphi(x)=\varphi(\tau,{\mathbf x})$ where
$x=(\tau,{\mathbf x}) \in {\mathbb R}^{(d+1)}$, $\tau \in {\mathbb R}$ and 
${\mathbf x} \in {\mathbb R}^{(d)}$.

The free Euclidean action, $S_0$, defined in terms of the phase-space 
variables is
\begin{equation}\label{eq:defs0}
S_0 \,=\,\int d^{d+1}x \,\Big[ - i \pi \partial_\tau \varphi 
+ {\mathcal H}_0(\pi,\varphi) \Big] \;,
\end{equation}
with
\begin{equation}
{\mathcal H}_0(\pi,\varphi) \;\equiv\; \frac{1}{2} \Big[ \pi^2 \,+\,
|{\mathbf \nabla}\varphi |^2 + m^2 \varphi^2 \Big] \;. 
\end{equation}

We then have to implement the periodic boundary conditions both for
$\varphi(\tau,{\mathbf x})$ and its canonical momentum $\pi(\tau,{\mathbf x})$
\begin{eqnarray}
\varphi\left(\beta,{\mathbf x}\right) &=& \varphi \left(0,{\mathbf
x}\right)\;,\;\;\forall\, {\mathbf x} \, \in {\mathbb R}^{(d)} \nonumber\\
\pi\left(\beta,{\mathbf x}\right) &=& \pi \left(0,{\mathbf
x}\right)\;,\;\;\forall\, {\mathbf x} \, \in {\mathbb R}^{(d)} \;,
\end{eqnarray}
which requires the introduction of two Lagrange multiplier fields,
as in the previous section about the harmonic oscillator. We denote them by
$\xi_a({\mathbf x})$, $a=1,\,2$  (they are $\tau$-independent).

To simplify matters, we now define a two-component field \mbox{$\Phi =
(\Phi_a)$}, \mbox{$a=1,\,2$}, such that $\Phi_1 = \varphi$ and $\Phi_2 = \pi$.
Then, an entirely analogous procedure to the one introduced for the harmonic
oscillator yields:
\begin{equation}\label{eq:zesc1}
{\mathcal Z}_0(\beta) \;=\; {\mathcal N} \,\int \,
{\mathcal D}\xi \, \int {\mathcal D}\Phi \; 
e^{-\frac{1}{2} \int d^{d+1}x \, \Phi_a 
{\hat{\mathcal K}}_{ab} \Phi_b \,+\,i\, \int d^{d+1}x  j_a \Phi_a}
\;,
\end{equation}
where:
\begin{equation}\label{eq:defja}
j_a(x) \equiv \xi_a({\mathbf x}) \, \big( \delta(\tau-\beta) 
- \delta(\tau) \big)
\;.
\end{equation}

Note that the measure for $\xi$ is formally given by:
\begin{equation}
{\mathcal D}\xi \;\equiv\; \prod_{\mathbf x} \frac{ d\xi_1({\mathbf x})
d\xi_2({\mathbf x})}{(2\pi)^2} \;,
\end{equation}
while:
\begin{equation}
{\mathcal D}\Phi \;\equiv\; \prod_{\tau,\mathbf x} 
\frac{ d\varphi(\tau,{\mathbf x}) d\pi(\tau, {\mathbf x})}{2\pi} \;,
\end{equation}
and the operator matrix ${\widehat{\mathcal K}}$ is:
\begin{equation}\label{eq:newdefk}
{\widehat{\mathcal K}}\;=\; 
\left( 
\begin{array}{cc}
{\hat h}^2       & i \frac{\partial}{\partial\tau} \\
- i \frac{\partial}{\partial\tau}     & 1 
\end{array}
\right) \;,
\end{equation}
where we have introduced \mbox{${\hat h} \equiv \sqrt{-\nabla^2 +
m^2}$}, the first-quantized energy operator for massive scalar 
particles.

Performing the integral over $\Phi$, and fixing ${\mathcal N}$ to verify
the normalization condition, yields the partition function in terms of the
Lagrange multipliers:
\begin{equation}
{\mathcal Z}_0(\beta) \;=\; 
\int {\mathcal D}\xi \,e^{- \frac{1}{2} \int d^dx
\int d^dy \, \xi_a ({\mathbf x}) \;
\langle {\mathbf x} | {\hat M}_{ab}| {\mathbf y} \rangle \; \xi_b ({\mathbf y})}
\;,
\end{equation}
with:
\begin{equation}\label{eq:newdefm}
{\hat M} \;\equiv\; 2 \, {\hat\Omega}(0)
\,-\, {\hat \Omega}(\beta) \,-\, 
{\hat\Omega}(-\beta)\;, 
\end{equation}
and
\begin{equation}\label{eq:newdefom}
{\hat \Omega}(\tau)\;\equiv\;
\left( 
\begin{array}{cc}
\frac{1}{2} {\hat h}^{-1} 
& \frac{i}{2} {\rm sgn}(\tau) \\
-\frac{i}{2} {\rm sgn}(\tau) 
& \frac{1}{2} {\hat h}
\end{array}
\right) \; e^{-{\hat h} |\tau|}  \;.
\end{equation}
Then,
\begin{equation}\label{eq:newm}
{\hat M}\;\equiv\;
\left( 
\begin{array}{cc}
{\hat h}^{-1} 
& 0 \\
0 
& {\hat h}
\end{array}
\right) \;({\hat n}_B + 1)^{-1}\;,
\end{equation}
where
\begin{equation}
{\hat n}_B \;\equiv\; \frac{1}{e^{\beta {\hat h}} - 1} \;.
\end{equation}
Coming back to the expression for ${\mathcal Z}_0(\beta)$, we see that:
\begin{eqnarray}\label{eq:zbose0}
{\mathcal Z}_0(\beta) &=& 
\int {\mathcal D}\xi \,\exp \Big\{ 
- \frac{1}{2} \int d^dx \int d^dy \big[ \xi_1 ({\mathbf x}) \;
\langle {\mathbf x} |{\hat h}^{-1} \, ({\hat n}_B+1)^{-1} | {\mathbf y} \rangle \;
\xi_1
({\mathbf y}) \nonumber\\
&+& \xi_2({\mathbf x}) \;\langle {\mathbf x} |{\hat h} \, ({\hat n}_B+1)^{-1} |
{\mathbf y} \rangle \; 
\xi_2 ({\mathbf y}) \big] \Big\} \;.
\end{eqnarray}
By a simple field redefinition, we see that:
\begin{equation}
{\mathcal Z}_0(\beta) \;=\; \det \big( {\hat n}_B + 1\big)
\end{equation}
which can be evaluated in the basis of eigenstates of momentum to yield:
\begin{equation}
{\mathcal Z}_0(\beta) \;=\; \prod_{\mathbf k} \big[ n_B(E_{\mathbf k}) + 1 \big]
\end{equation}
where $E_{\mathbf k} \equiv \sqrt{{\mathbf k}^2 + m^2}$. The 
free-energy density, $F_0(\beta)$, is of course:
\begin{equation}
F_0(\beta)\;=\; \frac{1}{\beta} \, \int \frac{d^dk}{(2\pi)^d} \, \ln \big( 1
\,-\, e^{-\beta E_{\mathbf k}}\big) \;.
\end{equation}

For the sake of completeness, we derive here the explicit form of the path
integral for the classical, high-temperature limit. When $\beta \sim 0$,
the integral for the partition function becomes:
\begin{equation}
{\mathcal Z}_0(\beta) \;\simeq\;
\int {\mathcal D}\xi \, e^{- \beta \; H(\xi)} \;,
\end{equation}
where:
\begin{equation}
H(\xi) \;=\;\frac{1}{2} \int d^dx \big[ \xi_1^2({\mathbf x}) \,+\,
|{\mathbf\nabla}\xi_2({\mathbf x})|^2 \,+\, m^2 \, \xi_2^2({\mathbf x}) 
\Big] \;.
\end{equation}
This is, again, the usual classical expression for the partition function,
with the Lagrange multipliers playing the role of phase space variables,
and the integration measure being the corresponding Liouville measure.
Besides, there is another fact that becomes here more transparent than in the
harmonic oscillator case, and we would like to point out. It is that the
representation (\ref{eq:zbose0}) always involves static fields, unlike in the
Matsubara formalism. The price to pay for this `dimensional reduction' is that
the resulting `action' (the exponent of the functional to be integrated) is
spatially non local. It becomes local only in the high-temperature limit.

\subsection{Self-interacting real scalar field}\label{ssec:self}
When field self-interactions are included, instead of the free action
$S_0$, we must consider instead
\begin{equation}
S\;=\;S_0 \,+\, S_I \;,
\end{equation}
where the free action $S_0$, has already been defined in (\ref{eq:defs0}), 
while $S_{I}$ is a self-interaction term. We shall assume it to be of the 
type:
\begin{equation}
S_I \;=\; \int d^{d+1}x \,V(\varphi) \;,
\end{equation}
$V(\varphi)$ being an even polynomial in $\varphi$, with only one (trivial)
minimum. Proceeding along similar lines to the ones followed for the free
field case in the preceding section, the partition function for the
interacting system can be written in the form:
\begin{equation}\label{eq:zescint1}
{\mathcal Z}(\beta) \;=\; {\mathcal N} \,\int \,
{\mathcal D}\xi \, \int {\mathcal D}\Phi \; 
e^{- S (\Phi) \,+\,i\, \int d^{d+1}x  j_a \Phi_a}
\;,
\end{equation}
where $\Phi$, as well as the `current' $j_a$ have already been defined for the
free case, in the previous subsection. The constant ${\mathcal N}$ is again
introduced to satisfy \mbox{${\mathcal Z}(\infty)=1$}. 
On the other hand, since the fields are assumed to tend to zero at infinity,
$\beta \to \infty$ implies that the term involving $j$ vanishes in this limit. 
This means that 
\begin{equation}
{\mathcal N}^{-1} \;=\; \,\int \,
{\mathcal D}\xi \, \int {\mathcal D}\Phi \; 
e^{- S (\Phi)}\;.
\end{equation}

There are many different paths one could follow from now on in order to
evaluate the partition function. We choose to adopt a procedure that makes
contact with quantities defined for QFT at $T=0$, in such a way that
the \mbox{$T \neq 0$} theory is built `on top of it'.

Indeed, recalling the definition of the generating functional for
connected correlation functions, ${\mathcal W}$, we may write:
\begin{equation}\label{eq:defw}
{\mathcal N} \, \int {\cal D}\Phi\; \exp\left[- S(\Phi) + i \int \,
d^{d+1}x j_a \Phi_a \right]\; \equiv \; e^{-{\mathcal W}(j)}\,,
\end{equation}
so that  the partition function ${\mathcal Z}(\beta)$ becomes:
\begin{equation}\label{eq:Wxi}
{\mathcal Z} (\beta) \;=\; \int {\cal D}\xi\; 
e^{- {\mathcal W}(j)}\;.
\end{equation}
We use the small $j$ to denote the $2$-component current which is a
function of the Lagrange multipliers, as defined in (\ref{eq:defja}). A
capital $J$ shall be reserved to denote a completely arbitrary
$2$-component source, so that:
\begin{equation}\label{eq:defw1}
{\mathcal N} \, \int {\cal D}\Phi\; exp\left[- S(\Phi) + i \int \,
d^{d+1}x J_a \Phi_a \right]\; \equiv \; e^{-{\mathcal W(J)}}\,.
\end{equation}

Note that this generating functional yields correlation functions including $\varphi$
or $\pi$ legs, so it would seem to contain much more information than the
standard one. That is not so, however, since
\mbox{${\mathcal W}(J_1, J_2)$} is related to \mbox{${\mathcal
W}(J_1)\equiv {\mathcal W}(J_1,0)$}, the standard generating functional of
$T=0$ $\varphi$-field correlation functions:
\begin{equation}
{\mathcal W}(J_1)\;\equiv\; - \ln 
\Big\{
\int {\mathcal D}\, \varphi \;e^{-\int d^{d+1}x \big[\frac{1}{2} (\partial
\varphi)^2 +\frac{1}{2} m^2 \varphi^2 + V(\varphi) - i J_1 \varphi 
\big]}\Big\}\;.
\end{equation}
That relation that may be obtained, for example, by integrating out the
canonical momentum field $\pi$ (a Gaussian integral):
\begin{equation}
{\mathcal W}(J_1,J_2) \,=\,{\mathcal W}(J_1\,-\,i \partial_\tau J_2)\;+\; 
\frac{1}{2} \int d^{d+1}x \, J_2^2(x) \;.  
\end{equation}
 The outcome of the previous relation is that correlation functions including
$\pi$ legs may be obtained by combining correlation functions where all the
canonical momenta have been replaced by scalar fields (at the same
spacetime points), and derivatives of them.

Besides, taking into account the fact that our normalization for ${\mathcal Z}(\beta)$
implies that ${\mathcal W}(0,0) = 0$, we are ready to write a formula for the
path-integral representation we were looking for. Indeed, defining
$H_{eff}(\xi)$, the `effective Hamiltonian' for ${\xi}$, by means of the 
expression:
\begin{equation}
H_{eff}(\xi) \;=\; \frac{1}{\beta} \, {\mathcal W}(j) \;,
\end{equation}
we see that the partition function is given by:
\begin{equation}
{\mathcal Z}(\beta)\; = \; \int {\mathcal D}\xi \; \exp\left[- \beta
H_{eff}(\xi) \right] \;.
\end{equation}
This is one of the most important results in this article, since it yields the
path integral for the quantum partition function as a classical-looking
functional integral, involving an effective Hamiltonian which takes into
account all the $T=0$ quantum effects. Indeed, the usual
functional expansion for ${\mathcal W}(J)$ is:
\begin{eqnarray}\label{eq:Wexpansion}
{\mathcal W}(J) &=& \sum_{n=2}^{\infty}\frac{1}{n!}\int d^{d+1}x_1\cdots
d^{d+1}x_n\, \nonumber\\
&\times& {\mathcal W}^{(n)}_{a_1 \ldots
a_n}(x_1,\,\ldots,\,x_n)\,J_{a_1}(x_1)\cdots J_{a_n}(x_n),
\end{eqnarray}
where ${\mathcal W}^{(n)}$ is $i^n$ times the $n$-point connected correlation
function for the field $\Phi$. Thus, knowing those correlation functions, at a
given order in a loop expansion, one may obtain an analogous functional
expansion for the effective Hamiltonian. Indeed, to do that, one should
perform the integrations over the imaginary-time variables (taking advantage of
the $\delta$-functions):
\begin{eqnarray}\label{eq:Heffexpansion}
H_{eff}(\xi) &=& \sum_{n=2}^{\infty}\frac{1}{n!}\int
d^d{\mathbf x}_1 \cdots d^d{\mathbf x}_n\,H^{(n)}_{a_1 \ldots
a_n}({\mathbf x}_1,\,\ldots,\,{\mathbf x}_n)\nonumber\\
&\times& \xi_{a_1}({\mathbf x}_1) \cdots \xi_{a_n}({\mathbf x}_n) \;,
\end{eqnarray}
where
\begin{eqnarray}
H^{(n)}_{a_1 \ldots a_n}({\mathbf x}_1,\,\ldots,\,{\mathbf x}_n) & \equiv &
\frac{1}{\beta} \, \int d\tau_1 \ldots d\tau_n \;
{\mathcal W}^{(n)}_{a_1 \ldots a_n}(x_1,\,\ldots,\,x_n) \nonumber\\
\times \; \big( \delta(\tau_1-\beta) - \delta(\tau_1) \big)
&\ldots& \big( \delta(\tau_n-\beta) - \delta(\tau_n) \big) \nonumber \;.
\end{eqnarray}

Of course, one will usually evaluate just a few of the terms in the expansion for
${\mathcal W}$, and as a consequence will obtain an approximation to the
effective Hamiltonian.
As an exercise, we evaluate the partition function that results from the first 
non-trivial approximation, namely, to keep only the first term, the
$2$-point correlation function~\footnote{The meaning of this approximation
shall be discussed at the end of this calculation.}:
\begin{equation}
H_{eff}(\xi) \sim  H^{(2)}_{eff}(\xi) 
\end{equation}
where
\begin{equation}
H^{(2)}_{eff}(\xi) \;\equiv\; \frac{1}{2} \, \int d{\mathbf x}_1 d{\mathbf x}_2
\xi_a({\mathbf x}_1)\, H^{(2)}_{ab}({\mathbf x}_1,\,{\mathbf x}_2)
\, \xi_b({\mathbf x}_2)
\end{equation}
and
\begin{eqnarray}\label{eq:Heffquad}
H^{(2)}_{ab}({\mathbf x}_1,\,{\mathbf x}_2) &=& \frac{1}{\beta}
\,\left[{\mathcal W}^{(2)}_{ab}(0,{\mathbf x}_1\,;0,
{\mathbf x}_2)+ {\mathcal W}^{(2)}_{ab}(\beta ,{\mathbf x}_1\,;\beta ,
{\mathbf x}_2)\right. \nonumber\\
& & \left.  - {\mathcal W}^{(2)}_{ab}(\beta ,{\mathbf x}_1\,;0,{\mathbf x}_2)-
{\mathcal W}^{(2)}_{ab}(0,{\mathbf x}_1\,;\beta ,{\mathbf x}_2)
\right] \;.
\end{eqnarray}
Introducing  Fourier transforms for the correlation functions
$\widetilde{\mathcal W}_{ab}$ and the kernel, $\widetilde{H}^{(2)}_{ab}$, 
we see that (\ref{eq:Heffquad}) implies: 
\begin{eqnarray}\label{eq:Heffquad1}
\widetilde{H}^{(2)}_{ab}(\omega,{\mathbf k}) &=& \frac{2}{\beta}
\,\int_{-\infty}^{+\infty} \frac{d\omega}{2\pi}\; 
\big[ 1 \,-\, \cos(\beta \omega) \big] 
\widetilde{\mathcal W}^{(2)}_{ab}(\omega,{\mathbf k})\;.
\end{eqnarray}

On the other hand, the kernels ${\mathcal W}^{(2)}_{ab}$ can be obtained from the
$T=0$ correlation function ${\mathcal W}^{(2)}_{11}$ which is, essentially,
the full quantum propagator for the $\varphi$-field:
\begin{equation}
{\mathcal W}^{(2)}_{11}(x_1,x_2) \;=\; -\, \langle \varphi (x_1) \varphi(x_2) \rangle \;.
\end{equation}
Indeed, the explicit relations between the  correlation functions
${\mathcal W}^{(2)}_{ab}$ with $a \neq 1$ or $b \neq 1$ and 
${\mathcal W}^{(2)}_{11}$, in momentum space, are:
\begin{eqnarray}
\widetilde{\mathcal W}^{(2)}_{22}(\omega,{\mathbf k}) 
&=& \,1\,-\,\omega^2 \widetilde{\mathcal W}^{(2)}_{11}(\omega,{\mathbf k}) 
\nonumber\\
\widetilde{\mathcal W}^{(2)}_{12}(\omega,{\mathbf k})&=& - \omega \, \widetilde{\mathcal
W}^{(2)}_{11}(\omega,{\mathbf k}) \nonumber\\
\widetilde{\mathcal W}^{(2)}_{21}(\omega,{\mathbf k}) &=&  \omega \widetilde{\mathcal
W}^{(2)}_{11}(\omega,{\mathbf k})) \;.
\end{eqnarray}
At this point, in order to evaluate the frequency integrals, we need of
course to make some assumptions about the structure of $\widetilde{\mathcal
W}^{(2)}_{11}$.  We shall assume that there is only one stable particle,
and that the renormalization conditions corresponding to the physical mass
and wavefunction renormalization have been imposed. Then 
$\widetilde{\mathcal W}^{(2)}_{11}$ has only two single poles,
at the points $\pm \, i \, E_{ren}({\mathbf k})$, where  $E_{ren}({\mathbf k})$ is the
dispersion relation, with a renormalized mass, and including all
the quantum corrections.  

Considering each component of equation (\ref{eq:Heffquad1}) for specific
values for $a$ and $b$, we see that their respective $\omega$ integrals are
easily evaluated under the previous assumptions. The results are:
\begin{equation}\label{eq:Heffquad11}
\widetilde{H}^{(2)}_{11}({\mathbf k}) \;=\; E^{-1}_{ren}({\mathbf k}) \, 
\Big[n_B\big(E_{ren}({\mathbf k})\big) + 1 \Big]^{-1} \;,
\end{equation}
\begin{equation}\label{eq:Heffquad22}
\widetilde{H}^{(2)}_{22}({\mathbf k}) \;=\; E_{ren}({\mathbf k}) \, 
\Big[n_B\big(E_{ren}({\mathbf k})\big)+1 \Big]^{-1}\;,
\end{equation}
while \mbox{$\widetilde{H}^{(2)}_{12}=\widetilde{H}^{(2)}_{21} =0$}.
 
Since the functional integral has the same structure as the one for the
free case, to evaluate ${\mathcal Z}(\beta)$ in this approximation we just need 
to modify the expression for the energy in the free field result:
\begin{equation}
{\mathcal Z}(\beta) \;\sim\; {\mathcal Z}^{(2)}(\beta)
\end{equation}
where
\begin{equation}
{\mathcal Z}^{(2)}(\beta) \;=\; \prod_{\mathbf k} \big[ n_B(E_{ren}({\mathbf
k})) + 1\big]
\end{equation}
which yields the free energy
\begin{equation}
F^{(2)}(\beta)\;=\; \frac{1}{\beta} \, \int \frac{d^dk}{(2\pi)^d} \, \ln \big( 1
\,-\, e^{-\beta E_{ren}({\mathbf k})}\big) \;.
\end{equation}
Of course, this expression looks like a the free energy for a free field,
but with the renormalized energies instead of the free ones.  This
approximation amounts, of course, to considering the particle states as non
interacting, after the self-energy corrections have been taken into account
(in the $2$-point, $T=0$ correlation function). 
A possible way to understand this approximation is by the large-$N$ limit,
whereby the correlation functions involving more than $2$ points are
suppressed, and the approximation used here becomes exact.

An interesting approximation scheme, related to the functional expansion, is
given by the High-Temperature expansion. In terms of $H_{eff}$, we see that
the only place where $\beta$ appears is in the source $j$. Performing an
expansion of $j$ in powers of $\beta$:
\begin{equation}
j_a(x) \;=\; \xi_a({\mathbf x}) \;\sum_{n=0}^\infty \,(-1)^n \, \frac{\beta^n}{n!}  
\delta^{(n)}(\tau) \;,  
\end{equation} 
and inserting this expansion into the expression for $H_{eff}$, we see that
the kernels corresponding to its functional expansion become:
\begin{eqnarray}
H_{a_1 \cdots a_m}^{(m)}({\mathbf x}_1,\cdots,{\mathbf x}_m) &=& \sum_{n_1=1, \cdots,n_m}^{\infty}
\frac{\beta^{n_1+ \cdots + n_m -1}}{n_1! \cdots n_m!} \nonumber\\
&\times & \big[\partial_{\tau_1} \cdots \partial_{\tau_m} {\mathcal W}^{(m)}_{a_1 \cdots a_m} 
(x_1,\cdots,x_m)\big]_{\tau_i \to 0} \;.
\end{eqnarray}

We see from this expression that, when $\beta \to 0$,  the leading term is
linear in $\beta$ and only $H^{(2)}$ survives:
\begin{equation}
H_{a_1 a_2}^{(2)}({\mathbf x}_1,{\mathbf x}_2) \simeq 
\beta^2 \big[\partial_{\tau_1}\partial_{\tau_2} {\mathcal W}^{(2)}_{a_1 a_2} 
(x_1,x_2)\big]_{\tau_1, \tau_2 \to 0} \;.
\end{equation}

We conclude this section mentioning the use of higher order
approximations to the effective Hamiltonian, say, by keeping $N$ terms in
the expansion:
\begin{eqnarray}\label{eq:HeffexpansionN}
H_{eff}^{(N)}(\xi) &=& \sum_{n=2}^{N}\frac{1}{n!}\int
d^d{\mathbf x}_1 \cdots d^d{\mathbf x}_n\,H^{(n)}_{a_1 \ldots
a_n}({\mathbf x}_1,\,\ldots,\,{\mathbf x}_n)\nonumber\\
&\times& \xi_{a_1}({\mathbf x}_1) \cdots \xi_{a_n}({\mathbf x}_n) \;,
\end{eqnarray}
(where only even values on $n$ are non-vanishing in the sum).
The resulting partition function naturally admits a perturbative expansion,
with the quadratic kernel $H^{(2)}_{ab}$ determining the propagator, and the
higher order terms defining nonlocal vertices.
\section{Dirac field}\label{sec:Dirac}
We shall derive here the effective Hamiltonian for the partition function
corresponding to a massive Dirac field in $d+1$ spacetime dimensions. The
procedure will be essentially the same as for the real scalar field, once
the relevant kinematical differences are taken into account.

As in the previous section, we shall first deal with the free field case.
\subsection{Free Theory}
The action $S_0^f$ for a free Dirac field $\psi$ in $d+1$ Euclidean dimensions is:
\begin{equation}\label{eq:defsD}
S_0^f \,=\,\int d^{d+1}x \; \bar{\psi}(x)(\not\!\partial + m)\psi(x)\;,
\end{equation}
where 
$\not\!\partial= \gamma_{\mu}\partial_{\mu}$, $\gamma_{\mu}^\dagger=\gamma_{\mu}$ 
and $\{\gamma_{\mu},\gamma_{\nu}\}=2 \delta_{\mu\nu}$.

The corresponding $T=0$ partition function, 
${\mathcal Z}_0^f(\beta)|_{\beta\rightarrow\infty}$, is then given by:
\begin{equation}\label{eq:defzif}
{\mathcal Z}_0^f(\infty) \;=\; \int \, 
{\mathcal D}\psi{\mathcal D}\bar{\psi}\, e^{-S_0^f(\bar{\psi},\psi)},
\end{equation}
where $\psi$,$\bar{\psi}$ are Grassmann-valued spinorial fields.

To derive the thermal partition function, ${\mathcal Z}_0^f(\beta)$, 
we must impose antiperiodicity conditions for both fields:
\begin{eqnarray}
\psi\left(\beta,{\mathbf x}\right) &=& -{\psi} 
\left(0,{\mathbf x}\right)\nonumber\\
\bar{\psi}\left(\beta,{\mathbf x}\right) &=& -\bar{\psi} 
\left(0,{\mathbf x}\right)
\end{eqnarray}
as constraints on the Grassmann fields in the path integral
(\ref{eq:defzif}).  Those conditions lead to the introduction of the two
$\delta-$functions:
\begin{eqnarray}
{\mathcal Z}_0^f(\beta) &=& \int \, {\mathcal D}\psi{\mathcal D}\bar{\psi}\,
\delta \big(\psi(\beta,{\mathbf x})+\psi(0,{\mathbf x})\big)\;
\delta \big(\bar{\psi}(\beta,{\mathbf x})
+\bar{\psi}(0,{\mathbf x})\big) \nonumber\\
&\times& \exp \Big[-S_0^f(\bar{\psi},\psi)\Big] \;.
\end{eqnarray}
Note that, since the Dirac action is of the first-order, the introduction
of two constraints, and two Lagrange multipliers, appears in an even more natural
way than for the previous case.  Those auxiliary fields, denoted by 
$\chi(\mathbf x)$ and $\bar\chi(\mathbf x)$
must be Grassmann spinors depending on the spatial coordinates only. 
The resulting expression for ${\mathcal Z}_0^f(\beta)$ 
can then be written as follows:
\begin{equation}
{\mathcal Z}_0^f(\beta) \;=\; {\mathcal N}\int \,  
{\mathcal D}\chi{\mathcal D}\bar{\chi}{\mathcal D}\psi{\mathcal D}\bar{\psi}\, 
e^{-S_0^f(\bar{\psi},\psi)+ i \, \int d^{d+1}x \,(\bar{\eta}\psi 
+ \bar{\psi}\eta )},
\end{equation}
where $\eta$ and $\bar{\eta}$ are (Grassmann) sources depending on 
$\chi$ and $\bar{\chi}$ through the relations:
\begin{eqnarray}
\eta(x)&=&\chi({\mathbf x})\big[\delta(\tau -\beta)
+\delta(\tau)\big]\nonumber\\
\bar{\eta}(x)&=&\bar{\chi} ({\mathbf x})\big[\delta(\tau -\beta)+\delta(\tau)\big]\;,
\end{eqnarray}
and the constant $\mathcal N$ is determined by imposing 
the usual normalization condition: \mbox{${\mathcal Z}_0^f(\infty)\equiv 1$}.

Integrating out $\psi, \bar{\psi}$:
$$
{\mathcal Z}_0^f(\beta) \;=\; \int \,  {\mathcal D}\chi{\mathcal D}\bar{\chi}\, 
\exp{\Big[ - \int d^{d+1}x \,\int d^{d+1}y \, 
\bar{\eta}(x)\langle x|(\not\!\partial + m)^{-1}|y\rangle \eta(y)\Big]}
$$
\begin{equation}
\;=\; \int \,  {\mathcal D}\chi{\mathcal D}\bar{\chi}\, 
\exp \Big[-\beta H_{eff}\big(\bar{\chi},\chi \big)\Big]
\end{equation}
where we have set ${\mathcal N}^{-1}\equiv {\mathcal Z}_0^f(\infty)$ and
\begin{equation}
H_{eff}\big(\bar{\chi},\chi \big)= \int d^{d}x \,\int d^{d}y \,
\bar{\chi}({\mathbf x})H^{(2)}\big({\mathbf x},{\mathbf y}\big )\chi({\mathbf y})
\end{equation}
with: 
\begin{eqnarray}
 H^{(2)}\big({\mathbf x},{\mathbf y}\big )&=& \langle \mathbf x,0|
 (\not\!\partial + m)^{-1}|\mathbf y,0 \rangle + \langle\mathbf x,\beta|(\not\!\partial 
 + m)^{-1}|\mathbf y,\beta \rangle \nonumber\\ 
 &+& \langle \mathbf x,0|(\not\!\partial + m)^{-1}|\mathbf y,\beta \rangle
 + \langle \mathbf x,\beta|(\not\!\partial + m)^{-1}|\mathbf y,0 \rangle\nonumber\\
 &=&\frac{1}{\beta}\,\Big[ 2\,S_f\big(0,\mathbf x - \mathbf y\big) 
 + S_f\big(\beta,\mathbf x - \mathbf y\big) 
 + S_f\big(-\beta,\mathbf x - \mathbf y\big)\Big].
\end{eqnarray}
In the last line, $S_f$, denotes the Dirac propagator.
Then a quite straightforward calculation shows that 
\begin{equation}
H\big({\mathbf x},{\mathbf y}\big )=\frac{1}{\beta}\,
\langle\mathbf x|\hat{u}(1 - \hat{n}_F)^{-1}|\mathbf y>
\end{equation}
where
\begin{equation}
\hat{n}_F \equiv \Big( 1+ e^{\beta \hat{n}}\Big)^{-1}
\end{equation}
is the Fermi-Dirac distribution function, written in terms of
\mbox{$\hat{h}$}, the energy operator (defined identically to its real
scalar field counterpart).  $\hat{u}$ is a unitary operator, defined as
\begin{equation}
\hat{u} \;\equiv\; \frac{\hat{h}_D}{\hat{h}}\, , \;\; \hat{h}_D \equiv
{\mathbf\gamma}\,\cdot\,{\mathbf \nabla} + m \;.
\end{equation}  

Then we easily verify that:
\begin{equation}
{\mathcal Z}_0^f(\beta) \;=\;\det\hat{u}\;
{\det}^{-1}\big[ (1-\hat{n}_F)\,\mathbf I\big] \;,
\end{equation}
($\mathbf I\,\equiv$ identity matrix in the 
representation of Dirac's algebra)
\begin{equation}
{\mathcal Z}_0^f(\beta) \;=\;\left\{\prod_{\vec{p}} 
\Big[ 1 + e^{-\beta E(\vec{p})}\Big] \right\}^{r_d}
\end{equation}
with $E({\mathbf p})=\sqrt{{\mathbf p}^2 + m^2}$ and $r_d\,\equiv$ 
dimension of the representation. 

We have used the fact that the determinant of $\hat{u}$ equals unity:
\begin{equation}
\det \hat{u} \; = \; \prod_{{\mathbf p}} \det 
\Big[ \frac{i{\mathbf \gamma}\cdot{\mathbf p} + m}{\sqrt{{\mathbf p}^2 + m^2}}
\Big]\,=\,1 \;. 
\end{equation}

Again, the procedure has produced the right result for the partition
function.
\subsection{Dirac field in a static external background}
We begin by highlighting the main differences which appear when an 
external, static (we deal with equilibrium thermal QFT), 
Abelian gauge field is coupled to the Dirac field.
The first is that the action becomes: 
\begin{equation}\label{eq:defsDA}
S^f (\bar{\psi},\psi, A)\;=\;\int d^{d+1}x \,
\Big[ \bar{\psi}(x)\big(\not\!\partial + i\,e\, 
{\mathbf\gamma}\cdot {\mathbf A} ({\mathbf x}) + m \big)\psi(x) \Big] \;,
\end{equation}
where we have used the $A_{0}=0$ gauge. The assumed $\tau-$ independence 
of ${\mathbf A}$, together with our gauge choice, allows us to 
carry on the derivation described for the free case, with minor changes, 
arriving to the expression:
\begin{eqnarray}
{\mathcal Z}^f(\beta) \;&=&\;\det\hat{u}({\mathbf A})\, 
{\det}^{-1}\big(\hat{n}_{F}({\mathbf A})\,\mathbf I\big)\nonumber\\
&=& e^{i K({\mathbf A})} \det\Big[\big( 1 
+ e^{-\beta \hat{h}({\mathbf A}) }\big)\mathbf I\Big]
\end{eqnarray}
where 
\begin{equation}
\hat{h}(A)\equiv \sqrt{-{\mathbf D}^2+m^2}\;,\;\;
{\mathbf D}\equiv {\mathbf\nabla}-i e {\mathbf A} \;,
\end{equation}
and:
\begin{equation}
e^{i K({\mathbf A})}\;=\;\frac{\det\big( {\mathbf\gamma}\cdot {\mathbf D} + m \big)}{\det
\sqrt{-{\mathbf D}^{2} + m^{2}}}\;.
\end{equation}
Notice that the factor \mbox{$\det\Big[\big( 1 + e^{-\beta \hat{n}({\mathbf A})}\big)
\mathbf I\Big]$}
can be formally diagonalized in terms of the energies $E_\lambda({\mathbf
A})$ in the presence of the external field. Thus we arrive to the
expression:  
\begin{equation}
{\mathcal Z}^f(\beta) \;=\;e^{i K({\mathbf A})}\;\times \;
\left\{\prod_\lambda \Big[ 1 + e^{-\beta E_\lambda({\mathbf A})}\Big] 
\right\}^{r_d} \;.
\end{equation}
The factor $e^{i K({\mathbf A})}$, on the other hand, is topological in origin, as it depends on 
the phase, $K({\mathbf A})$, of the determinant of $\hat{h}_D$, an operator that may be regarded as a Dirac operator in one fewer dimension.
For Dirac fermions, we know that the phase of $\det\hat{h}_D$ can be non-trivial only when
$d$ is odd, i.e., when $d+1$ is even. However, the $\gamma$-matrices appearing
in $\det\hat{h}_D$ form a reducible representation of the Dirac algebra in $d$ dimensions, 
with the matrix $\gamma_\tau$ relating every eigenvalue to its complex
conjugate. Thus, as a result, the phase $K({\mathbf A})$ vanishes. Of
course, a non-vanishing result may be obtained for other fermionic systems,
like Weyl fermions for \mbox{$d+1={\rm even}$}.

\newpage
\section{Conclusions}\label{sec:conclu}
We have shown that, by introducing the `thermal' boundary conditions 
as constraints in the Euclidean path integral for the vacuum functional,
we may obtain a novel representation for the partition function. 
This representation may be thought of as an integral over the
phase space variables, weighted by a Boltzmann factor corresponding to an
effective quantum Hamiltonian, $H_{eff}$, which reduces to the classical
one in the corresponding (high-temperature) limit.  

We analyzed the main properties of this representation for the cases of the
real scalar and Dirac fields, two typical examples that have been chosen
for the sake of simplicity. It is not difficult to generalize the
representation to the case of systems containing  fermions interacting with
bosons. For example, assuming that $S({\bar\psi},\psi,\Phi)$ is the first
order action corresponding to a real scalar interacting with a Dirac field,
we define the $T=0$ generating functional \mbox{${\mathcal W}({\bar
\zeta},\zeta, J)$} by:
\begin{equation}
{\mathcal W}({\bar \zeta},\zeta, J) \;=\; {\mathcal N}\int \,  
{\mathcal D}\Phi {\mathcal D}\psi{\mathcal D}\bar{\psi}\, 
e^{-S(\bar{\psi},\psi,\Phi)+ i \,\int d^{d+1}x \,\big( \bar{\zeta}\psi + \bar{\psi}\zeta  
+ J_a \Phi_a \big)},
\end{equation}
where the sources are arbitrary. Then, $H_{eff}$ can be obtained from the
expression:
\begin{equation}
H_{eff}({\bar\chi},\chi,\xi) \;=\; \frac{1}{\beta} \;{\mathcal W}({\bar \eta},\eta, j) \;, 
\end{equation}
where $\eta$, ${\bar \eta}$ and $J$  are (the already defined) functions of
the Lagrange multiplier fields ${\bar\chi}$, $\chi$ and $\xi$.

We conclude by mentioning that $\tau$-dependent correlation functions can also be calculated, with minor changes in the formalism. Indeed, they may be obtained by adding an arbitrary time-dependent source to the current that depends on the constraints~\cite{fmr}.

\section*{Acknowledgments}
C.D.F. thanks CONICET (PIP5531) and CAPES/SPU for financial support, and acknowledges the warm hospitality of the CBPF, where part of this work was done. A.P.C.M. and I.R. thank CAPES/SPU, CNPq/MCT and Pronex/FAPERJ for partial financial support 
and Instituto Balseiro - CAB, where part of this work has been done, for kind hospitality.

\end{document}